\documentstyle[aps,prb,epsfig,float,twocolumn]{revtex}
\begin{document}
\twocolumn[\hsize\textwidth\columnwidth\hsize\csname
@twocolumnfalse\endcsname

\title{Dissipation effects in spin-Hall transport of electrons and holes}

\author{John Schliemann and Daniel Loss}

\address{Department of Physics and Astronomy, University of Basel,
CH-4056 Basel, Switzerland}

\date{\today}

\maketitle

\begin{abstract}
We investigate the spin-Hall effect of both electrons and holes in 
semiconductors using the Kubo formula in the correct 
zero-frequency limit taking into
account the finite momentum relaxation time of carriers in real
semiconductors. This approach allows to analyze the range of validity of 
recent theoretical findings. In particular, the spin-Hall conductivity 
vanishes for vanishing spin-orbit coupling if the correct zero-frequency 
limit is performed.
\end{abstract}
\vskip2pc]

\section{Introduction}

In the recent years, an increasing interest in spin-dependent phenomena in
semiconductors has developed, mostly in the field of spin electronics,
which has by now become a major branch of solid state research
\cite{Wolf01,Awschalom02,Rashba03}. 
On of the most investigated issues in this field is the influence of 
spin-orbit coupling on various transport properties of both electrons and 
holes. Many of these studies were inspired by the paradigmatic proposal
of a spin field-effect transistor due to Datta and Das \cite{Datta90};
for recent work in this direction see e.g. Refs.~\cite{Schliemann03a,Pala02,Saikin02,Mishchenko03,Lusakowski03,Winkler03,Ganichev03,Schliemann03b,Hall03}.
Most recently, interesting theoretical studies
on the spin-Hall effect have been performed 
\cite{Murakami03,Sinova03,Culcer03}. This effect amounts in a spin
current (as opposed to a charge current) driven by an electric field 
perpendicular to it. In the present letter we reexamine these findings using
the Kubo formula with full frequency dependence, treating both the case
of electrons \cite{Sinova03} and holes \cite{Murakami03,Culcer03},
and analyze the range of validity of previous theoretical results
obtained for the case of direct current. Here it is crucial
to perform the correct zero-frequency limit taking into account an
imaginary part of the frequencies occurring the the Kubo formula.

The notion of the spin-Hall effect in systems of itinerant spinful carriers
was considered first by Dyakonov and Perel \cite{Dyakonov71} 
in the early seventies, and, 
independently, in a more recent paper by Hirsch \cite{Hirsch99}.
In these studies the predicted spin-Hall effect is due to spin-orbit
effects influencing scattering processes upon static impurities.
Following the terminology used in \cite{Sinova03,Culcer03} this is referred
to as the {\em extrinsic} spin-Hall effect since it depends on impurity
scattering. This is in contrast to the {\em intrinsic}
spin-Hall effect predicted very recently in Refs. 
\cite{Murakami03,Sinova03,Culcer03} which is entirely
due to spin-orbit coupling terms in the single-particle carrier Hamiltonian
and independent of any scattering process. As we shall see below, this
distinction between intrinsic and extrinsic effects becomes ambiguous in
the limit of weak spin-orbit coupling when life time effects 
of carrier quasiparticles have to be taken into account.

Yet another type of spin-Hall effect was studied recently by Meier and Loss
\cite{Meier03}
in a two-dimensional Heisenberg model consisting of isolated spins,
in contrast to the itinerant-carrier systems mentioned before.

Spin-orbit coupling also induces off-diagonal components of the conductivity 
tensor for charge transport. An important example is the anomalous Hall
effect as it occurs in semiconductors in the presence of magnetic 
impurities \cite{Jungwirth02}. Here, as in the case of the aforementioned
spin-Hall effect, the off-diagonal elements
of the charge conductivity tensor are the same in magnitude but differ in sign.
Therefore, this antisymmetric 
conductivity tensor has the same components in all
orthogonal coordinate systems, and in this sense the transport properties
are isotropic. This is different from charge transport of electrons in
quantum wells as investigated recently Ref.~\cite{Schliemann03b}.
In such systems, the presence of spin-orbit coupling of both 
the Rashba \cite{Rashba60} and the Dresselhaus \cite{Dresselhaus55}
type leads to anisotropic dispersion relations and Fermi contours.
This feature leads to symmetric off-diagonal elements in the
conductivity tensor and therefore to preferred eigendirections for
charge transport \cite{Schliemann03b,Ganichev03}. This predicted effect
offers a possibilty to detect spin-orbit coupling by measuring diffussive
spin-unpolarized charge currents in a Hall-type geometry, which should be a
comparatively simple experimental task.

This paper is organized as follows. In section \ref{Kubo} we summarize
elementary properties of linear-response theory as given by the Kubo formula.
This technique is applied then in section \ref{electrons} to spin-Hall
transport of electrons in a quantum well in the presence of Rashba spin-orbit
coupling. In section \ref{holes} we investigate the case of bulk holes
described by the Luttinger Hamiltonian in the spherical approximation.
We end with conclusions in section \ref{conclusions}.

\section{Kubo formula and zero-frequency limit}
\label{Kubo}

Our present study of spin-Hall effect of electrons and holes in semiconductors
is based on the usual Kubo formula with full frequency dependence for a
spatially homogeneous electric field \cite{Mahan00},
\begin{eqnarray}
\sigma^{S,z}_{xy}(\omega) & = & \frac{e}{A(\omega+i\eta)}
\int_{0}^{\infty}e^{i(\omega+i\eta)t}\nonumber\\
 & & \cdot\sum_{\vec k,\mu}f(\varepsilon_{\mu}(\vec k))
\langle\vec k,\mu|[j^{S,z}_{x}(t),v_{y}(0)]|\vec k,\mu\rangle\,,
\label{generalKubo}
\end{eqnarray}
where we have assumed zero temperature $T=0$ and non-interacting carriers,
which allows to formulate the two-body Green's function entering the
conductivity Kubo formula in terms of single-particle operators.
$A$ is the volume of the system, $e$ is the elementary charge, and
$f(\varepsilon_{\mu}(\vec k))$ is the $T=0$ Fermi distribution function 
for  energy $\varepsilon_{\mu}(\vec k)$ at wave vector $\vec k$ in a
dispersion branch labeled by $\mu$.
The velocity operators are given by $\vec v=i[{\cal H},\vec r]/\hbar$
where $\vec r$ is the position operator, and ${\cal H}$ is the 
single-particle Hamiltonian 
not including the external electric field. The spin-current operator
(in the Dirac picture)
for spin moment polarized along the $z$-direction and flowing in the
$x$-direction is given by 
\begin{equation}
j^{S,z}_{x}(t)=
e^{i{\cal H}t/\hbar}\frac{1}{2}\left(s^{z}v_{x}+v_{x}s^{z}\right)
e^{-i{\cal H}t/\hbar}\,,
\end{equation}
where $\vec s$ is the spin operator.
The right hand side of Eq.~(\ref{generalKubo}) has to be understood
in the limit of vanishing imaginary part $\eta>0$ in the frequency argument.
This imaginary part in the frequency reflects the
fact that the external electric field is assumed to be switched on 
adiabatically starting from the infinite past of the system, and it also
ensures causality properties of the retarded Green's function occurring in
Eq.~(\ref{generalKubo}). In general, and as we will discuss in more detail 
below, the limiting process $\eta\to 0$ does not commute with other
limits, and, in particular, the dc-limit $\omega\to 0$ has to be taken with 
care \cite{Mahan00}. In the presence of random impurity scattering,
the retarded two-body Green's function in Eq.~(\ref{generalKubo}) will 
generically have
a frequency argument with positive imaginary part \cite{Mahan00}. 
In this case the limit 
$\eta\to 0$ is unproblematic, and the imaginary part of the
frequency argument is just due to impurity scattering and/or other
(many-body) effects. Generically, and as we will discuss in more detail 
below, the imaginary part $\eta>0$ corresponds to a finite carrier 
quasiparticle lifetime.

\section{Electrons with Rashba coupling} 
\label{electrons}

Sinova {\em et al.} \cite{Sinova03}
have considered the spin-Hall effect of non-interacting electrons
confined to the two-dimensional ($xy$-) plane of a quantum well and 
being subject to Rashba spin-orbit coupling \cite{Rashba60,Rashba03},
\begin{equation}
{\cal H}=\frac{\vec p^{2}}{2m^{*}}+\frac{\alpha}{\hbar}
\left(p_{x}\sigma^{y}-p_{y}\sigma^{x}\right)\,,
\label{Rashba}
\end{equation}
where $m^{*}$ is the effective mass, $\alpha$ the Rashba coefficient, and
the other notations are standard. We note that in systems where both
the Rashba and the Dresselhaus spin-orbit coupling 
\cite{Dresselhaus55,Rashba03} are present, various 
interesting transport effects can arise from the interplay of these two terms,
for recent studies see e.g. \cite{Schliemann03a,Saikin02,Mishchenko03,Lusakowski03,Ganichev03,Schliemann03b}. For simplicity, however, we shall concentrate 
here on the Rashba term only. The Hamiltonian (\ref{Rashba}) has
two energy branches,
\begin{equation}
\varepsilon_{\pm}(\vec k)=\frac{\hbar^{2}k^{2}}{2m^{*}}\pm\alpha k
\end{equation}
with eigenstates
\begin{equation}
\langle\vec r|\vec k,\pm\rangle
=\frac{e^{i\vec k\vec r}}{\sqrt{A}}\frac{1}{\sqrt{2}}
\left(\begin{array}{c}
 1 \\ \pm e^{i\chi(\vec k)}
\end{array}\right)\,,
\end{equation}
where $\chi(\vec k)=\arg(-k_{y}+ik_{x})$.
By a straightforward calculation one obtains for the spin-Hall
conductivity
\begin{eqnarray}
\sigma^{S,z}_{xy}(\omega) & = & -\sigma^{S,z}_{yx}(\omega)\nonumber\\
 & = & -\frac{e}{4\pi}\frac{\alpha}{m^{*}}
\int^{k_{f}^{-}}_{k_{f}^{+}}dk
\frac{k^{2}}{(\omega+i\eta)^{2}-\left(\frac{2\alpha k}{\hbar}\right)^{2}}\,,
\label{elecsigma1}
\end{eqnarray}
where
\begin{equation}
k_{f}^{\pm}=\sqrt{\frac{2m^{*}}{\hbar^{2}}\varepsilon_{f}
+\left(\frac{m^{*}}{\hbar^{2}}\right)^{2}\alpha^{2}}
\mp\frac{m^{*}}{\hbar^{2}}\alpha
\end{equation}
are the Fermi momenta on the two dispersion branches 
for positive Fermi energy $\varepsilon_{f}>0$.
In the presence of scattering on static random impurities, the
imaginary part $\eta>0$ in the frequency argument 
is given, to lowest order in the Rashba coefficient and the
impurity potential, by the inverse of the momentum relaxation time.
This is certainly a very intuitive result; however, let us sketch a formal
proof for this assertion. The time-dependent spin-current
operator in the presence of Rashba coupling reads
\begin{equation}
j^{S,z}_{x}(t)=\frac{\hbar}{2m^{*}}\sigma^{z}(t)p_{x}(t)
\end{equation}
where the time
evolution includes impurity scattering. To lowest order in the
spin-orbit coupling and the impurity scattering we have
\begin{equation}
j^{S,z}_{x}(t)\approx \frac{\hbar}{2m^{*}}\sigma^{z}_{0}(t)p_{x}^{0}(t)
\end{equation}
where the time evolution of $\sigma^{z}_{0}$ is only due to the
Hamiltonian (\ref{Rashba}) and evaluated in the above expression
(\ref{elecsigma1}), while $p_{x}^{0}(t)$ contains
the impurity scattering but not the spin-orbit coupling.
Now it is useful to note that, in order to compute the
expectation values in the Kubo formula Eq.~(\ref{generalKubo}),
only matrix elements of the time-dependent momentum operator
$p_{x}^{0}(t)$ which are diagonal in the wave vector index are needed. 
This enables to apply superoperator
techniques developed in Refs.~\cite{Loss86} yielding
\begin{equation}
\left(p_{x}^{0}(t)\right)_{\vec k\vec k}\approx
\left(e^{-\Omega_{0}t}p_{x}^{0}(0)\right)_{\vec k\vec k}
=\left(e^{-t/\tau}p_{x}^{0}(0)\right)_{\vec k\vec k}\,,
\end{equation}
where $\Omega_{0}$ is the scattering master operator in lowest 
order of the scattering potential \cite{Loss86}. It is the same
operator as it occurs as the scattering term in the usual Boltzmann
equation when evaluated in lowest oder via Fermi's golden rule.
For impurity potentials being isotropic in real space, the momentum 
$p_{x}$ is an exact eigenfunction of $\Omega_{0}$, and the 
eigenvalue is given by the well-known inverse momentum relaxation 
time $1/\tau(\varepsilon)$ 
\cite{Loss86,Smith89} which in general depends on the energy
$\varepsilon(\vec k)$. To lowest order in the Rashba coupling,
this energy argument can be replaced with the Fermi energy in the absence of
spin-orbit interaction. We note that this momentum relaxation rate
$1/\tau$ is the same as obtained in the standard diagrammatic approach
and thus contains the vertex correction \cite{Mahan00}. However, this
vertex correction vanishes for short-range isotropic scatterers.
The above argumentation refers to the
Rashba Hamiltonian (\ref{Rashba}) for conduction band electrons; 
similar considerations can be performed in the case of valence  band holes 
to be discussed further below.

For $\omega=0$, but finite momentum relaxation rate $1/\tau>0$, 
Eq.~(\ref{elecsigma1}) yields
\begin{eqnarray}
 & & \sigma^{S,z}_{xy}(0)=-\sigma^{S,z}_{yx}(0)
=\frac{e}{8\pi}
-\frac{e}{32\pi}\frac{\hbar/\tau}{\varepsilon_{R}}
\nonumber\\
 & & \cdot\tan^{-1}\left(4\frac{\varepsilon_{R}}{\hbar/\tau}
\left(1+8\frac{\varepsilon_{R}\varepsilon_{f}}
{(\hbar/\tau)^{2}}\right)^{-1}\right)\,,
\end{eqnarray}
where we have introduced the ``Rashba energy'' 
$\varepsilon_{R}=m^{*}\alpha^{2}/\hbar^{2}$. Clearly, this is the energy  
which has to be compared with the energy scale $\hbar/\tau$ 
of the impurity scattering in order to obtain the correct zero-frequency
limit of the spin Hall conductivity. If the impurity scattering is
weak compared to spin-orbit coupling, $\varepsilon_{R}\tau/\hbar\gg 1$,
we have the expansion
\begin{equation}
\sigma^{S,z}_{xy}(0)=\frac{e}{8\pi}-\frac{e}{64\pi}
\frac{(\hbar/\tau)^{2}}{\varepsilon_{R}\varepsilon_{f}}
+{\cal O}\left(\left(\frac{\hbar/\tau}{\varepsilon_{R}}\right)^{2}\right)
\label{expansion1}
\end{equation}
where we have additionally assumed that the Fermi energy $\varepsilon_{f}$
is larger or at least of the same order as $\varepsilon_{R}$, which
is usually the case in experimental situations.  
The zeroth-order contribution $e/8\pi$ 
is the result obtained by Sinova {\em et al.} using directly a 
zero-frequency perturbative expression for the spin-Hall current
neglecting effects of a finite electron quasi-particle lifetime,
(cf. Eq.~(8) in Ref.~\cite{Sinova03}). Remarkably, this value
is universal in the sense that it is independent
of $\alpha$. Therefore, it predicts a finite spin-Hall conductivity
even in the limit of vanishing spin-orbit coupling, $\alpha\to 0$, which is
certainly an unphysical feature. However, this paradox can be resolved
by the observation that the above two limiting processes do not
commute. In fact, in the opposite limit $\varepsilon_{R}\tau/\hbar\ll 1$
the lowest order of the second term on the r.h.s. of Eq.~(\ref{elecsigma1}) 
cancels the first one, and the spin-Hall conductivity is given in leading 
order by
\begin{equation}
\sigma^{S,z}_{xy}(0)=\frac{e}{\pi}
\frac{\varepsilon_{R}\varepsilon_{f}}{(\hbar/\tau)^{2}}
+{\cal O}\left(\left(\frac{\varepsilon_{R}}{\hbar/\tau}\right)^{2}\right)
\label{expansion2}
\end{equation}
Thus, to obtain the correct dc spin-Hall conductivity, the ``Rashba energy''
$\varepsilon_{R}=m^{*}\alpha^{2}/\hbar^{2}$ should be compared with the
energy scale $\hbar/\tau$ of the impurity scattering. If 
$\varepsilon_{R}\gg \hbar/\tau$ the spin-Hall conductivity is close to its
``universal'' value $e/8\pi$, while it vanishes for small spin-orbit
coupling and finite impurity scattering. In epitaxially grown GaAs 
quantum wells mobilities $\mu=e\tau/m^{*}$ of order $100{\rm m^{2}/Vs}$
can routinely be achieved, corresponding to values for
$\hbar/\tau$ of order $0.01{\rm meV}$. This is safely smaller
than typical values for the Rashba energy reported from experiments 
\cite{Nitta97,Engels97,Heida98,Hu99,Grundler00,Sato01,Lommer88,Jusserand92,Jusserand95}being of order $0.1\dots 1.0{\rm meV}$. 
However, it should be noted that the Rashba
coefficient is typically proportional to an external electric field 
applied in the growth direction of the quantum well. Therefore also
smaller values of the Rashba energy are possible where the finite
momentum relaxation time will influence the value of the spin-Hall
conductivity.

\section{Holes in the valence band of III-V semiconductors}
\label{holes}

Murakami, Nagaosa, and Zhang have investigated spin-Hall transport
in three-dimensional bulk systems of holes in the valence band of
III-V semiconductors \cite{Murakami03}. These authors used a
phenomenological semiclassical theory to describe adiabatic hole
dynamics. Their work was revisited most recently by Culcer {\em et al}.
\cite{Culcer03} within the framework of a semiclassical theory of wave packet 
dynamics. Here we will evaluate the spin-Hall conductivity using the
rigorous Kubo formula (\ref{generalKubo}). Our starting point 
\cite{Murakami03,Culcer03} is Luttinger's four-band Hamiltonian
for heavy and light holes in the spherical approximation 
\cite{Luttinger56},
\begin{equation}
{\cal H}=\frac{1}{2m}\left(\left(\gamma_{1}+\frac{5}{2}\gamma_{2}\right)
\vec p^{2}-2\gamma_{2}\left(\vec p\cdot\vec S\right)^{2}\right)\,.
\label{Luttinger}
\end{equation}
Here $m$ is the bare electron mass, and $\vec S$ are spin-$3/2$-operators.
The dimensionless Luttinger parameter $\gamma_{1}$ and $\gamma_{2}$ 
describe the valence
band of the specific material with effects of spin-orbit coupling being
included in $\gamma_{2}$. 
The eigenstates of (\ref{Luttinger})
can be chosen to be eigenstates of the helicity operator
$\lambda=(\vec k\cdot\vec S)/k$. The heavy holes correspond to
$\lambda=\pm 3/2$, while the light holes have $\lambda=\pm 1/2$.
From the Kubo formula (\ref{generalKubo}) one finds for the 
frequency-dependent spin-Hall conductivity after lengthy but
elementary calculations
\begin{eqnarray}
\sigma_{xy}^{S,z} (\omega)
& = & -\frac{e}{\pi^{2}}\left(\frac{\hbar}{m}\right)^{2}
\left(\gamma_{1}+2\gamma_{2}\right)\gamma_{2}\nonumber\\
 & & \cdot\int_{k_{f}^{l}}^{k_{f}^{h}}dk
\frac{k^{4}}{\left(\omega+i/\tau\right)^{2}
-\left(\frac{2\hbar}{m}\gamma_{2}k^{2}\right)^{2}}\,,
\label{holescond1}
\end{eqnarray}
where
\begin{equation}
k_{f}^{h/l}=\sqrt{\frac{2m}{\hbar^{2}}\varepsilon_{f}
\frac{1}{\gamma_{1}\mp2\gamma_{2}}}
\end{equation}
are the Fermi wave numbers for
heavy and light holes, respectively. Again it is instructive to consider
the case for weak spin-orbit coupling, $\gamma_{2}\ll\gamma_{1}$.
For $\gamma_{2}=0$ we have 
$k_{f}^{h}=k_{f}^{l}=:k_{F}^{0}=\sqrt{2m\varepsilon_{f}/\gamma_{1}\hbar^{2}}$,
and therefore the integral in Eq.~(\ref{holescond1}) vanishes
for finite $1/\tau>0$ and all frequencies $\omega$. Thus, as 
before for the case of electrons, the dc spin-Hall conductivity vanishes for
vanishing spin-orbit coupling if a finite momentum relaxation rate
is taken into account. This result is in contrast to statements in 
Refs.~\cite{Murakami03,Culcer03}, where such dissipation effects were
neglected. Specifically, for $\omega=0$ we have
\begin{eqnarray}
\sigma_{xy}^{S,z} (0)
 & = & \frac{e}{4\pi^{2}}\frac{\gamma_{1}+2\gamma_{2}}{\gamma_{2}}\nonumber\\
 & & \cdot\left(k_{f}^{h}-k_{f}^{l}
-\int_{k_{f}^{l}}^{k_{f}^{h}}dk
\frac{1}{1
+\left(\frac{2}{\hbar/\tau}\frac{\hbar^{2}}{m}\gamma_{2} k^{2}\right)^{2}}
\right)
\label{holescond2}
\end{eqnarray}
The remaining integral is elementary leading to a rather tedious
expression which shall not be given here.
However, we
see that the value of the above integral is governed by the ratio of
$\hbar/\tau$ and the ``spin-orbit energy''
$\varepsilon_{so}:=\hbar^{2}\gamma_{2}(k_{f}^{0})^{2}/m
=2\varepsilon_{f}\gamma_{2}/\gamma_{1}$, since $k_{f}^{0}$ is a typical
wave number in the integration interval. If 
$\hbar/\tau\gg \varepsilon_{so}$ the spin-Hall
conductivity vanishes as
\begin{eqnarray}
\sigma_{xy}^{S,z} (0) & = & 
\frac{e}{\pi^{2}}4k_{f}^{0}
\left(\frac{\varepsilon_{so}}{\hbar/\tau}\right)^{2}
\frac{\gamma_{2}}{\gamma_{1}}\nonumber\\
 & & +{\cal O}\left(\left(\frac{\varepsilon_{so}}{\hbar/\tau}\right)^{4}
,\left(\frac{\varepsilon_{so}}{\hbar/\tau}\right)^{2}
\left(\frac{\gamma_{2}}{\gamma_{1}}\right)^{2}\right)
\label{expansion3}
\end{eqnarray}
where we have also assumed that the ratio $\gamma_{2}/\gamma_{1}$ is small
as it is usually the case \cite{Vurgaftman01}.
In the opposite case $\hbar/\tau\ll \varepsilon_{so}$ one finds
\begin{eqnarray}
\sigma^{S,z}_{xy}(0)
 & = & \frac{e}{4\pi^{2}}\frac{\gamma_{1}+2\gamma_{2}}{\gamma_{2}}
\Biggl[k_{f}^{h}-k_{f}^{l}\nonumber\\
 & & +\frac{\left(k_{f}^{0}\right)^{4}}{12}
\left(\left(\frac{1}{k_{f}^{h}}\right)^{3}
-\left(\frac{1}{k_{f}^{l}}\right)^{3}\right)
\left(\frac{\hbar/\tau}{\varepsilon_{so}}\right)^{2}\Biggr]\nonumber\\
 & & +{\cal O}\left(\left(\frac{\hbar/\tau}{\varepsilon_{so}}\right)^{4}\right)
\label{expansion4}
\end{eqnarray}
We note that the zeroth order of this  result agrees with the
expression given in 
Ref.~\cite{Culcer03} for the dc spin-Hall conductivity 
neglecting dissipation effects \cite{note1}, but
differs from the result reported in \cite{Murakami03}. On the present stage we
cannot comment on the question whether this difference is an artifact
of the semiclassical approache used in Ref.~\cite{Murakami03},
whereas the present result is obtained from a rigorous linear-response
theory given by the Kubo formula with full frequency dependence.

Let us illustrate our results on the typical example given
in Ref.~\cite{Murakami03}, where a GaAs sample with hole density
$n=10^{19}{\rm cm^{-3}}$, corresponding to a Fermi energy of order a few
ten meV, and a mobility of $\mu=e\tau/m^{*}=50{\rm cm^{2}/Vs}$. To obtain an
upper bound for $\tau$ we take $m^{*}$ to be the heavy-hole mass,
$m^{*}\approx 0.5 m$, corresponding to $\gamma_{1}\approx 7$ and
$\gamma_{2}\approx 2.5$ for GaAs \cite{Vurgaftman01}. This leads to a lower
estimate for $\hbar/\tau$ being also of order a few ten meV. Thus, in the
above scenario, the finite momentum relaxation time $\tau$ must be taken
into account when calculating the spin-Hall conductivity, differently
from the approach in Ref.~\cite{Murakami03}.

\section{Conclusions}
\label{conclusions}
We have studied the spin-Hall transport of electrons and holes in 
semiconductors using the Kubo formula in the correct 
zero-frequency limit taking into
account the finite momentum relaxation time of carriers in real
semiconductors. This approach allows to analyze the range of validity of 
recent theoretical findings\cite{Murakami03,Sinova03,Culcer03}. 
In particular, the spin-Hall conductivity is found to 
vanishe for vanishing spin-orbit coupling if the correct zero-frequency 
limit is performed. In the case of conduction band electrons in the
presence of Rashba spin-orbit coupling in a high-mobility quantum well,
spin-orbit interaction dominates, for typical experimental
for the rashba coefficient, the effects of momentum relaxation, and the
spin-Hall conductivity is close to its ``universal'' value as predicted in
Ref.~\cite{Sinova03}. This situation can be different for typical
p-doped bulk samples, where dissipation can substantially affect the
spin-Hall transport.

\acknowledgments{We thank J. Sinova and A.~H. MacDonald for useful 
correspondence.
This work was supported by NCCR Nanoscience, the Swiss NSF,
DARPA, and ARO.}


\begin{references}
\bibitem{note1}
The zeroth order in Eq.~(\ref{expansion4}) agrees with the second expression
given in Eq.~(18) of Ref. \cite{Culcer03} for the spin-Hall conductivity,
up tp a factor of $1/3$ which is due to different definitions of the
spin-current operator.
\end{references}
\end{document}